\documentclass[letterpaper]{article} 
\usepackage{aaai24}  
\usepackage{times}  
\usepackage{helvet}  
\usepackage{courier}  
\usepackage[hyphens]{url}  
\usepackage{graphicx} 

\usepackage{multirow}
\usepackage{xcolor}
\usepackage{booktabs}
\usepackage{adjustbox}
\definecolor{lightthulianpink}{rgb}{0.9, 0.56, 0.67}

\usepackage{xcolor,colortbl}
\usepackage{natbib}  
\usepackage{caption} 
\usepackage{algorithm}
\usepackage{algorithmic}
\usepackage{amsmath}
\usepackage{amsthm}
\usepackage{amsfonts}
\usepackage{newfloat}
\usepackage{listings}
\DeclareCaptionStyle{ruled}{labelfont=normalfont,labelsep=colon,strut=off} 
\floatstyle{ruled}
\newfloat{listing}{tb}{lst}{}
\floatname{listing}{Listing}
\title{Leveraging Large Language Models for Pre-trained Recommender Systems}

\author{
    Zhixuan Chu\equalcontrib\textsuperscript{\rm 1},
    Hongyan Hao\equalcontrib\textsuperscript{\rm 1},
    Xin Ouyang\textsuperscript{\rm 1},
    Simeng Wang\textsuperscript{\rm 1},
    Yan Wang\textsuperscript{\rm 1},
    Yue Shen\textsuperscript{\rm 1},
    Jinjie Gu\textsuperscript{\rm 1},
    Qing Cui\textsuperscript{\rm 1},
    Longfei Li\textsuperscript{\rm 1},
    Siqiao Xue\textsuperscript{\rm 1},
    James Y Zhang\textsuperscript{\rm 1},
    Sheng Li\textsuperscript{\rm 2}
}
\affiliations{
    \textsuperscript{\rm 1}Ant Group\\
    \textsuperscript{\rm 2}University of Virginia\\
    \ \{chuzhixuan.czx, hongyanhao.hhy, xin.oyx, simeng.wsm, luli.wy, zhanying, jinjie.gujj,  cuiqing.cq, longyao.llf, siqiao.xsq, james.z\}@antgroup.com, shengli@virginia.edu 

}

\usepackage{bibentry}

\begin{document}

\maketitle

\begin{abstract}

Recent advancements in recommendation systems have shifted towards more comprehensive and personalized recommendations by utilizing large language models (LLM). However, effectively integrating LLM's commonsense knowledge and reasoning abilities into recommendation systems remains a challenging problem. In this paper, we propose RecSysLLM, a novel pre-trained recommendation model based on LLMs. RecSysLLM retains LLM reasoning and knowledge while integrating recommendation domain knowledge through unique designs of data, training, and inference. This allows RecSysLLM to leverage LLMs' capabilities for recommendation tasks in an efficient, unified framework. We demonstrate the effectiveness of RecSysLLM on benchmarks and real-world scenarios. RecSysLLM provides a promising approach to developing unified recommendation systems by fully exploiting the power of pre-trained language models.

\end{abstract}

\section{Introduction}

The realm of recommendation has gained considerable attention in recent years due to its ability to drive business growth and enhance user engagement. Recent advancements in recommender systems have shifted towards incorporating diverse information and catering to a broader range of application scenarios, rather than focusing on task-specific architectures. This shift has been driven by the need for more comprehensive and personalized recommendations, as well as the availability of new data sources and knowledge \cite{geng2022recommendation,chu2022hierarchical,hui2022personalized,sheu2021knowledge,li2021survey,jiang2022learning,xue2021graphpp}. In addition, with the advent of the Large Language Model (LLM) \cite{radford2018improving,radford2019language,brown2020language,ouyang2022training}, we have witnessed an unprecedented surge in the capabilities of natural language processing. The power of LLM lies in its ability to understand and generate human-like language. LLM has also enabled the extraction of implicit knowledge from text data \cite{gu2023robust,yoneda2023llm,zhao2023logic}. This newfound capability of LLM has opened up exciting avenues for the integration of semantic information into recommender systems and provides a wealth of insights into user preferences and behaviors \cite{shi2023language,zhao2022tiny}. As a result, incorporating LLM into recommender systems has become a crucial step toward providing a powerful and comprehensive paradigm for recommendation tasks. In the following, we will discuss the new generation of recommendation model paradigms from two directions, i.e., the unified pre-trained recommendation model and the combination of LLM and recommendation model.

On the one hand, training a pre-trained recommendation model can help overcome the limitations of existing recommendation approaches that require designing task-specific architectures and training objectives. Traditional recommendation methods have focused on a single task, such as personalized product recommendations, contextual advertising, customer segmentation, and so on, making them less adaptable to new tasks and limiting their ability to generalize to new domains. By training a pre-trained recommendation model, we can leverage the power of pre-trained models to learn generalizable representations of user behavior  and product characteristics \cite{tsai2023game,zhao2022tiny} that can be applied to a variety of recommendation tasks. Overall, a pre-trained recommendation model provides a flexible and scalable solution that can be adapted to a variety of recommendation tasks. Since recommendation tasks usually share a common user–item pool, features, behavioral sequences, and other contextual information, we believe it is promising to merge even more recommendation tasks into a unified framework so that they can implicitly transfer knowledge to benefit each other and enable generalization to other unseen tasks \cite{xie2022nc}. 

On the other hand, integrating LLMs into recommendation systems has several significant advantages. These advantages are linked to the LLM's capabilities in thinking, reasoning, and discovering implicit relationships within textual data based on the entailment of wealthy background knowledge and logical chains. (1) By leveraging the semantic information in natural language data, LLMs can help the recommendation system understand and infer the relationship between user features and behavioral sequences and among entities in behavioral sequences.  This allows the recommendation system to understand the user's needs and preferences in a more comprehensive way. (2) Another benefit of integrating LLMs into recommendation systems is the ability to leverage the implicit knowledge that is hidden in the models. LLMs are trained on vast amounts of textual data and can help to understand the relationships between different concepts and ideas. By incorporating LLMs into recommendation systems, this implicit knowledge can be used to generate more divergent and logical recommendations. This can lead to more creative and unexpected recommendations that the user may not have considered otherwise. (3) By leveraging the natural language processing capabilities of LLMs, recommendation tasks that previously required separate specialized systems can now be integrated into a unified framework. 
The pretrained knowledge and few-shot learning abilities of LLMs allow recommendation models to be rapidly adapted to new domains with limited data. Overall, the natural language processing power and versatility of LLMs can help merge more recommendation tasks into a unified framework. Furthermore, a comprehensive survey on recommendations and LLMs is provided in the Appendix. This survey covers the motivation behind them, current development, and challenges.

However, constructing a robust and integrated recommendation system that fully utilizes large language models' immense knowledge and reasoning capacities poses several key challenges. Directly training a pre-trained recommendation model from scratch is not only a waste of time and data collection efforts but also lacks general common sense and reasoning capabilities that underpin modern large language models. Meanwhile, directly fine-tuning a pre-trained LLM model on recommendation data also has drawbacks. Recommendation data has distinct characteristics - such as fixed entities and sequential user behaviors - that differ from the raw text corpora used to train language models. As such, fine-tuning may erase much of the capabilities specific to recommendation tasks. Therefore, we propose a novel pre-trained recommendation paradigm (RecSysLLM) based on the pre-trained large language model through unique designs for recommendation in three phases, i.e., data phase, training phase, and inference phase. Our model retains the reasoning ability and rich knowledge contained in large language models while integrating the recommendation-specific knowledge. It directly inherits the parameters and framework of the original large language model but also designs and extends some mechanisms in the data phase (textualization and sampling), training phase (mask, position, and ordering), and inference phase (dynamic position infilling). These modifications do not discard the tokenization, parameters, structure, or previously learned knowledge in the LLM. On this basis, recommendation data is used to fine-tune it. The significant advantage of this pre-trained recommendation model is that it can utilize the reasoning capabilities and rich knowledge of large language models while incorporating domain-specific knowledge of the recommendation system through parameter-efficient fine-tuning of user-profiles and behavioral sequences data. Another crucial benefit of this model is that it can be easily adapted to different downstream recommendation sub-tasks. We evaluate the proposed model on extensive benchmark datasets and real-world scenarios. The experimental results demonstrate its effectiveness in improving the quality of recommendations. Overall, our proposed pre-trained recommendation model provides a promising approach for building recommendation systems that are efficient, effective, and unified.

\section{RecSysLLM Pretraining Mechanism}

\begin{figure*}[t]
\centering
\includegraphics[width=1\textwidth]{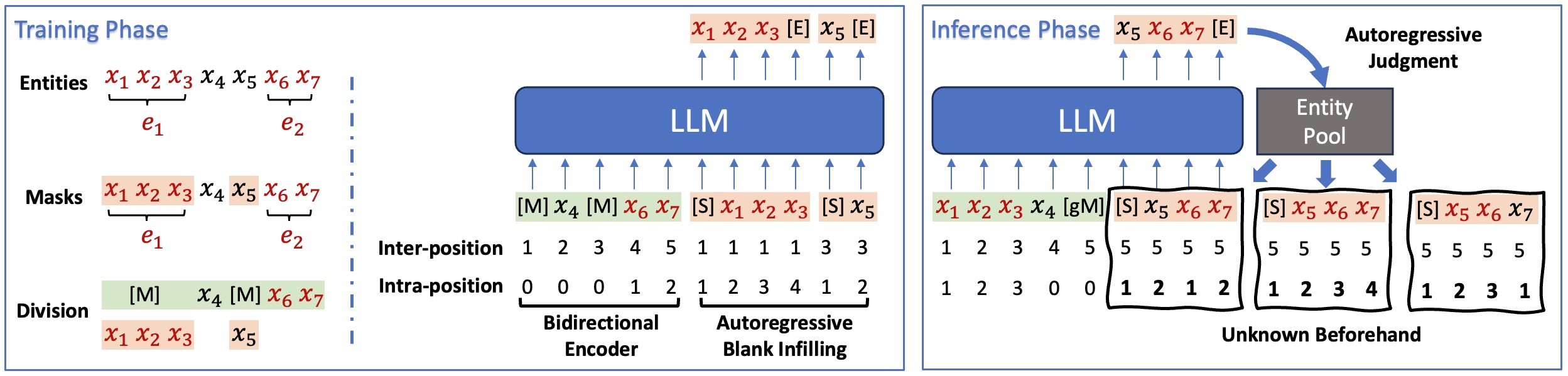}

\caption{This is the framework of RecSysLLM based on a pre-trained generative language model (GLM). To transform the GLM into a specialized model for recommendation systems, several modifications are made while preserving the core knowledge and capabilities of the original language model architecture, such as the new mask mechanism, span order, positional encoding, dynamic position mechanism, and so on.}

\label{RecSysLLM_framework}
\end{figure*}

To fully take advantage of LLM and domain knowledge in recommendation tasks, we need to modify the LLM and fine-tune the existing LLM to get a pre-trained recommendation model. However, the conventional large language models are trained on general knowledge and coherent corpus, and the framework of the model is not designed for behavioral sequence data and recommendation tasks. To address these two points, we make modifications from three phases, i.e., data, training, and inference phases, to transform a conventional pre-trained language model into a pre-trained recommendation model. The whole framework is illustrated in Figure \ref{RecSysLLM_framework}. This pre-trained recommendation model has been employed in real-world applications in Chinese scenarios, so we take the GLM \cite{du2021glm} as an example to introduce the RecSysLLM pretraining mechanism, which is bilingual in Chinese and English. Our model can also be adapted to other large language models with minor modifications.

\subsection{Data Phase}
In the data phase, textualizing tabular data is often the easiest and most straightforward approach for implementing large language models. For the pre-training of RecSysLLM, we first textualize conventional tabular data, such as user features stored in a table with rows and columns into text. Since large language models are originally trained on textual data, text-based features can be easily combined with text-based behavioral sequences and other text information, which helps our model better capture the relationship between features and behavioral sequences. In addition, textualizing tabular data allows for greater flexibility in how they are used in the following tasks. 

Compared with ordinary language texts, the training texts in the recommendation system should take into account the interests and preferences of users from different periods \cite{yu2019adaptive}. Long-term preferences are usually stable and reflect the general preferences of a user. These preferences do not change frequently over time, but they lack timeliness and may not reflect current interests. On the other hand, short-term preferences tend to change frequently over time and are more reflective of a user's current interests. We aim to use different periods of preferences to provide accurate and relevant recommendations to users, which can balance the user's general interests with their current needs. Therefore, we sample behavioral sequences in long-term preferences ($10\%$), medium-term preferences ($30\%$), and short-term preferences ($60\%$). Long-term preferences capture the user's preferences that have remained consistent for an extended period of time, typically spanning over several months or years. Medium-term preferences capture the user's preferences that have developed and changed over a shorter period of time, typically spanning over several weeks or months. Short-term preferences can improve recommendation accuracy by providing the system with the user's most recent preferences, spanning over several days or hours.

\subsection{Training Phase}

To be consistent with the architecture of GLM, our model is still trained by optimizing an autoregressive blank infilling objective based on an input text $\boldsymbol{x}=[x_1,\cdots,x_n]$. Different from the general language text in GLM, our input text is composed of user features and behavioral sequences. Although textualized user features and behavioral sequences are also composed of multiple tokens, they often represent a complete meaning as a whole. If they are split into different parts, like regular text, they will lose their unique meaning. In addition, the LLM's power comes from the way it tokenizes and processes text. It has been trained on a vast amount of data and has learned to recognize patterns and relationships between tokens, enabling it to identify entities accurately and extract information. If we were to create a new tokenization method, we would lose the LLM's power. Therefore, to maintain the LLM's power and supplement the new knowledge in the recommendation data, it is best to leverage the existing tokenization and enhance it with additional information and capabilities rather than create a new tokenization. In the following, we name the attributes in user features and items in the behavioral sequences as entities, which means that they are complete units and have fixed meanings. Therefore, as shown in the ``Entities'' of Figure \ref{RecSysLLM_framework}, our data are composed of plain language text and entities, where ($x_1$, $x_2$, and $x_3$) have merged to form $e_1$ and ($x_6$ and $x_7$) to form $e_2$. $x_4$ and $x_5$ are separate tokens.

\paragraph{Mask Mechanism.} To inject the new knowledge of recommendation tasks based on the original LLM, we follow the principle in the LLM and design the new mask mechanism and position strategies. Similar to the GLM \cite{du2021glm}, multiple text spans $\{\boldsymbol{s}_1,\cdots,\boldsymbol{s}_m\}$ are sampled, where each span $\boldsymbol{s}_i$ corresponds to a series of consecutive tokens $[s_{i,1},\cdots,s_{i,l_i}]$ in $\boldsymbol{x}$. Each span is replaced with a \textit{single} $[\text{MASK}]$ token. The remaining text and $[\text{MASK}]s$ form a corrupted text $\boldsymbol{x}_{\text{corrupt}}$. In the GLM, since there is no existence of \textit{entity}, the tokens can be randomly sampled into spans. However, in our model, the multiple and consecutive tokens composing an entity should not be split into different parts. In other words, the tokens of an entity are treated as a whole. The $[\text{MASK}]$ mechanism will not break the \textit{complete} entities, which will highlight the whole structure of entities and help to capture the interrelationship between entities. For example, as shown in the ``Masks'' of Figure \ref{RecSysLLM_framework}, $x_1$, $x_2$, and $x_3$ composing the $e_1$ are blocked as a whole and single token $x_5$ is also blocked. Therefore, we form the $\boldsymbol{x}_{\text{corrupt}}$ with $[\text{M}]$, $x_4$, $[\text{M}]$, $x_6$, and $x_7$ in the ``Division'' of Figure \ref{RecSysLLM_framework}.

Compatible with different natural language processing tasks, we adopt the multi-task pretraining setup \cite{du2021glm} with entity-level $[\text{M}]$, sentence-level $[\text{sM}]$, and document-level $[\text{gM}]$. Specifically, entity-level refers to the randomly blanking out continuous spans of tokens from the input text, following the idea of autoencoding, which captures the interdependencies between entities. Sentence level restricts that the masked spans must be full sentences. Document-level is to sample a single span whose length is sampled from a uniform distribution over 50\%--100\% of the original length. The objective aims for long text generation.

\paragraph{Span Order.} We implement the autoregressive blank infilling objective with the following techniques. The input $\boldsymbol{x}$ is divided into two parts: one part is the corrupted text $\boldsymbol{x}_{\text{corrupt}}$, and the other consists of the masked spans. Our model automatically learns a bidirectional encoder for the first part and a unidirectional decoder for the second part in a unified model. The model predicts the missing tokens in the spans from the corrupted text in an autoregressive manner, which means when predicting the missing tokens in a span, the model has access to the corrupted text and the previously predicted spans. Instead of randomly permuting the order of the spans in the original GLM \cite{du2021glm}, we keep all spans in chronological order to keep the interrelationship among different entities. Formally, we define the pretraining objective of a length-$m$ index sequence $[1,2,...,m]$ as
\begin{equation}
     \sum_{i=1}^m\log p(\boldsymbol{s}_{i}|\boldsymbol{x}_{\text{corrupt}},\boldsymbol{s}_1,...,\boldsymbol{s}_{i-1};\theta)
    \label{eqn:objective}
\end{equation}
\paragraph{Positional Encoding.}
To enable autoregressive generation, each span is padded with special tokens [START] and [END], for input and output, respectively. To be consistent with the original LLM, we cannot arbitrarily modify, add, or reduce the original positional strategies. Therefore, we extend 2D positional encodings \cite{du2021glm} based on entities. Specifically, each token is encoded with two positional ids, i.e., inter-position and intra-position ids. 

The inter-position id represents the position in the corrupted text $\boldsymbol{x}_{\text{corrupt}}$. For the masked spans, it is the position of the corresponding [MASK] token. For the intra-position id, we follow the essential meaning in the original LLM, which still refers to the intra-position. Instead of the scope of the whole span, we extend it into a finer granularity. For the entities, it represents the intra-relationship among entities. As shown in Figure \ref{RecSysLLM_framework}, for separate tokens (not in the entities) in the encoder part ($[\text{M}]$, $x_4$, $[\text{M}]$), their intra-position ids are 0. For consecutive tokens in the entities ($x_6$ and $x_7$), they are numbered in chronological order. For tokens in the autoregressive blank infilling part, they range from 1 to the length of the entities including [S], such as (entities: [S], $x_1$, $x_2$, $x_3$ $\rightarrow$ $1,2,3,4$) and (independent token: [S], $x_5$ $\rightarrow$ $1,2$ ). The two positional ids are projected into two vectors via learnable embedding tables, which are both added to the input token embeddings.

\begin{figure}[th!]
\centering
\includegraphics[width=0.9\columnwidth]{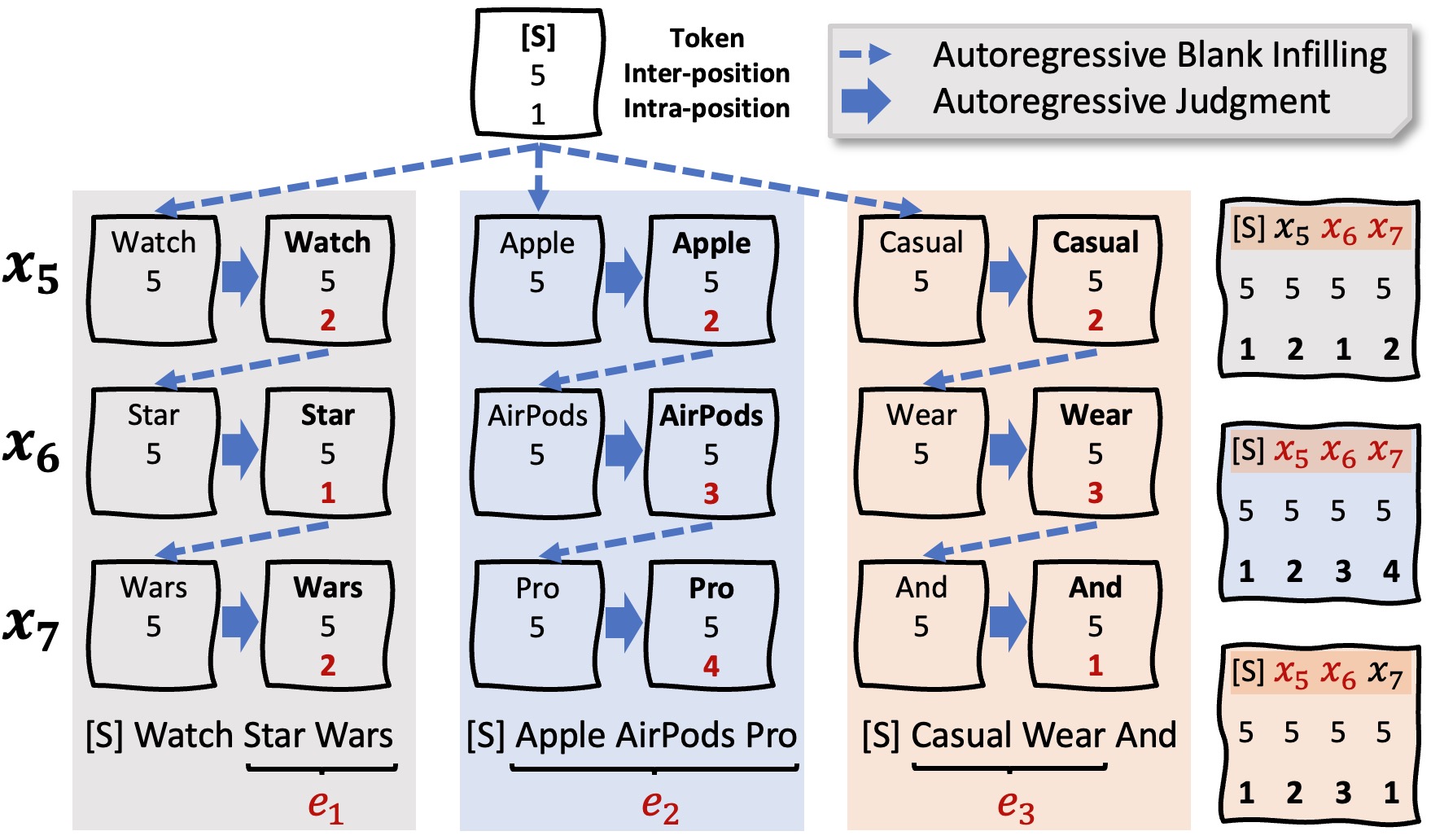}

\caption{This is the dynamic position mechanism. When one token is generated, it will be judged as one part of an entity or not. If it and the previous token belong to one entity, the intra-position id will continue to grow. Otherwise, it will start at $1$ again.}

\label{RecSysLLM_example}
\end{figure}

\subsection{Inference phase}
Because our pre-trained model is designed to fit different downstream tasks, the length of the generated text should be unknown beforehand and flexible for the different tasks. Further, due to the existence of entities, the intra-position ids represent the relative position of the entity. As shown in the ``Inference Phase'' of Figure \ref{RecSysLLM_framework}, we cannot specify the intra-position ids in advance when autoregressive blank infilling. Hence, we designed a dynamic position mechanism for the mask and position modifications made during the inference phase. It can conduct the autoregressive judgment to determine and complement the intra-position ids one by one as each token is generated in the autoregressive generation procedure. Specifically, we establish an entity pool beforehand, which stores all the tokens of the entities that existed in our recommendation task. When one token is generated, it will be judged as one part of an entity or not. We utilize the Trie algorithm \cite{bodon2003trie} to check whether the generated token and previous token belong to the same entity, which is a tree data structure used for locating specific keys from within a set. If they belong to one entity, the intra-position id will continue to grow. Otherwise, it will start at $1$ again. The detailed procedure is illustrated in Figure \ref{RecSysLLM_example}.

\section{Experiments}

\subsection{Experimental Setup} 
\label{sec:setup}
\paragraph{Datasets.} 
We evaluate our method on three real-world e-commerce datasets from Amazon.com, spanning the categories of Sports \& Outdoors, Beauty, and Toys \& Games. The datasets contain user ratings and reviews from 2019, along with transaction records between January 1 and December 31 \cite{zhou2020s3,xue2022hypro,xue2023easytpp}. Key statistics of the resulting datasets are provided in Table~\ref{tab:stats}.

\paragraph{Metrics.} Following the experiments in \cite{geng2022recommendation}, we cover five different task families – rating, sequential recommendation, explanation, review, and direct recommendation to facilitate the multitask pretraining for the recommendation. For rating prediction, we adopt Root Mean Square Error (RMSE) and Mean Absolute Error (MAE) as evaluation metrics. For sequential recommendation and direct recommendation tasks, we employ top-$k$ Hit Ratio (HR@$k$) and Normalized Discounted Cumulative Gain (NDCG@$k$) to evaluate the performance and report HR@{1, 5, 10} and NGCG@{5, 10}. For explanation generation and review summarization, we evaluate different methods with BLEU-4, ROUGE-1, ROUGE-2, and ROUGE-L. Lower values of RMSE and MAE indicate better performance, while higher values are preferred for all other metrics. In all result tables, \textbf{bold} numbers represent the best performance, while \underline{underlined} numbers refer to the second-best performance.

\paragraph{Baselines for Multiple Tasks}

To demonstrate competence on a wide range of recommendation-related tasks, we adopt the same representative approaches as \cite{geng2022recommendation} for different tasks, such as Rating Prediction (\textit{MF}~\cite{koren2009matrix} and \textit{MLP}~\cite{cheng2016widedeep}), Direct Recommendation (\textit{BPR-MF} \cite{rendle2009bpr}, \textit{BPR-MLP} \cite{cheng2016widedeep}, and \textit{SimpleX} \cite{mao2021simplex}), Sequential Recommendation (\textit{Caser}~\cite{tang2018personalized}, \textit{HGN}~\cite{ma2019hierarchical}, \textit{GRU4Rec}~\cite{hidasi2016gru4rec}, 
\textit{BERT4Rec}~\cite{sun2019bert4rec}, \textit{FDSA}~\cite{zhang2019feature}, \textit{SASRec}~\cite{kang2018self}, and \textit{S$^3$-Rec}~\cite{zhou2020s3}), Explanation Generation (\textit{Attn2Seq}~\cite{EACL17-Att2Seq}, 
\textit{NRT}~\cite{li2017neural},
\textit{PETER}~\cite{li2021personalized}, and \textit{PETER+}), and review summarization (\textit{T0}~\cite{sanh2022multitask} and \textit{GPT-2}~\cite{radford2019language}). The detailed baselines are provided in the Appendix.

\begin{table}[t!]
\centering
\small

\caption{Basic statistics of the experimental datasets.}

\resizebox{0.65\columnwidth}{!}{
\begin{tabular}{lrrrr}
\toprule
Dataset &  {\textbf{Sports}} &  {\textbf{Beauty}} & {\textbf{Toys}} \\
\cmidrule{1-4}
\#Users    &  35,598   &  22,363   &  19,412  \\
\#Items    &  18,357   &  12,101  & 11,924  \\
\#Reviews  & 296,337   &  198,502   & 167,597 \\
\#Sparsity (\%) &  0.0453  & 0.0734  &  0.0724 \\
\bottomrule
\end{tabular}}

\label{tab:stats}
\end{table}

\subsection{Implementation}
\label{implementation}

To facilitate the multitask prompt-based pretraining for the recommendation, \citet{geng2022recommendation} created a collection of personalized prompt templates. The collection covers five different task families – rating, sequential recommendation, explanation, review, and direct recommendation. The prompts include personalized fields for users and items to help the model discover user-item preferences. For rating prediction, prompts ask to predict a user's rating or preference for an item. For sequential recommendation, prompts ask to predict the next item a user will interact with. For explanation, prompts ask to generate text explaining a user's preferences. For review, prompts summarize or predict ratings from reviews. For direct recommendation, prompts ask whether to recommend an item to a user. The complete collection of personalized prompts with examples is provided in the Appendix of \cite{geng2022recommendation}. These prompts enable the building of diverse training examples from raw data for multitask pertaining. We pretrain our RecSysLLM with diverse training examples with different prompt templates from all five task families to verify its multitask learning ability. Besides, we adopt a part of prompts in each task family for zero-shot evaluation while all remaining prompts are utilized for multitasking prompted pretraining. As a result, we are able to not only compare the performance across various recommendation tasks but also evaluate the zero-shot generalization capability on unseen prompts.

Our RecSysLLM model for these English language tasks leverages the powerful GLM-10B for English \cite{du2021glm} model as a foundation. GLM is a General Language Model pretrained with an autoregressive blank-filling objective and can be finetuned on various natural language understanding and generation tasks. Our approach builds on this pre-trained GLM-10B foundation by utilizing a parameter-efficient fine-tuning method called LoRA (Low-Rank Adaptation) \cite{hu2021lora} to adapt the model to our specific recommendation tasks. LoRA enables efficiently customizing the enormous GLM-10B model to specialized domains by learning a low-dimensional decomposition of the model update. This allows us to tap into GLM-10B's broad language knowledge while calibrating it to our RecSysLLM objectives. We inject trainable rank decomposition matrices into each $query\_key\_value$, $dense$, $dense\_h\_to\_4h$ and $dense\_4h\_to\_h$ layer of Transformer architecture in GLM-10B.
We pretrain our RecSysLLM for eight epochs with AdamW optimization \cite{loshchilov2017decoupled} on four NVIDIA RTX A100 GPUs. In order to achieve efficient use of memory and distributed training, we use the DeepSpeed \cite{rasley2020deepspeed} module. The batch size is set to 32 per GPU. We set the peak learning rate as $1 \times 10^{-5}$  and use a warmup strategy to adjust the learning rate. In addition, we set the maximum length of input tokens to 1024.

\subsection{Performance.} 
\begin{table}[t!]
\centering
\footnotesize

\caption{Performance on rating prediction. The shadow refers to the test on unseen prompts in a zero-shot manner.}

\resizebox{0.8\columnwidth}{!}{
\begin{tabular}{ccccccc}
\toprule
\multirow{2.5}{*}{Methods} & \multicolumn{2}{c}{\textbf{Sports}} & \multicolumn{2}{c}{\textbf{Beauty}} &  \multicolumn{2}{c}{\textbf{Toys}} \\
\cmidrule(lr){2-3}\cmidrule(lr){4-5}\cmidrule(lr){6-7}
 & RMSE  & MAE & RMSE  & MAE & RMSE  & MAE \\
\cmidrule{1-7}
MF    &  \bf 1.0234  & 0.7935  & \bf 1.1973 & 0.9461 & \underline{1.0123} & 0.7984    \\
MLP  & 1.1277 & 0.7626  & 1.3078  & 0.9597  & 1.1215 & 0.8097  \\
P5 {\color{purple}}   &  1.0357  & \underline{0.6813}  & 1.2843 & 0.8534 & 1.0544  & 0.7177    \\
RecSysLLM {\color{purple}}   &  1.0410  & 0.7012  & 1.2721 & \underline{0.8431} & 1.0246  & 0.7012    \\
\rowcolor{lightthulianpink} P5 {\color{orange}}  &  1.0292  & 0.6864  &  1.2870 & 0.8531 & 1.0245 &  \underline{0.6931}   \\
\rowcolor{lightthulianpink} RecSysLLM {\color{orange}}  &  \underline{1.0278}  & \bf 0.6631  &  \underline{1.2671} & \bf 0.8235 & \bf 1.0112 &  \bf 0.6014   \\
\bottomrule
\end{tabular}}

\label{tab:rating}
\end{table}

\begin{table*}[t!]
\centering

\caption{The training sequences in Amazon Toys dataset for P5 and our RecSysLLM model.}

\resizebox{2\columnwidth}{!}{
\begin{tabular}{c|ll}
\toprule
 Sequence  & P5 & RecSysLLM\\
 \midrule
 1 &1, 2, 3, 4, 5, 6, 7  & Hasbro Electronic Catch Phrase,\quad  Gloom,\quad Cards Against Humanity,\quad Carcassonne Basic Game,\quad\\
 && Asmodee 7 Wonders Wonder Pack,\quad Village Board Game,\quad Rory's Story Cubes - Voyages\\

 \midrule
 2 &8, 9, 10, 11, 12 & Megabloks CAT 3in1 Ride On Truck,\quad Fisher-Price Jake and The Never Land Pirates - Jake's Musical Pirate Ship Bucky,\quad\\
 && VTech KidiBeats Drum Set,\quad Playskool Heroes Transformers Rescue Bots Blades the Copter-Bot Figure,\quad LeapFrog LeapPad2 Power Learning Tablet\\
\midrule
 
 1767 &692, 5235, 5765, 709, 7162& Badger Basket White Doll Crib With Cabinet Bedding And Mobile - Pink/White,\quad Badger Basket Doll High Chair With Plate Bib And Spoon - Pink/White,\quad\\
 &&  Fisher-Price Brilliant Basics Lil Snoopy (Colors May Vary),\quad LeapFrog Shapes and Sharing Picnic Basket,\quad JC Toys 20\&quot; La Baby Doll\\

 \midrule
 17788 &10092, 9958, 8925, 2881, 2706 & The Walking Dead TV Board Game,\quad Zombie Survival Playing Cards,\quad McFarlane Toys The Walking Dead Comic Series 2 Penny The Governors Daughter Action Figure,\quad\\
 &&  Webkinz Velvety Elephant,\quad Webkinz Love Frog Limited Edition Release\\

\bottomrule
\end{tabular}}

\label{training_example}
\end{table*}

\begin{table*}[ht!]
\centering
\caption{Performance on the sequential recommendation. The shadow refers to the test on unseen prompts in a zero-shot manner.}

\resizebox{1.9\columnwidth}{!}{
\begin{tabular}{ccccccccccccc}
\toprule
\multirow{2.5}{*}{Methods} & \multicolumn{4}{c}{\textbf{Sports}} & \multicolumn{4}{c}{\textbf{Beauty}} &  \multicolumn{4}{c}{\textbf{Toys}} \\
\cmidrule(lr){2-5}\cmidrule(lr){6-9}\cmidrule(lr){10-13}
 & HR@5  & NDCG@5 & HR@10  & NDCG@10 & HR@5  & NDCG@5 & HR@10  & NDCG@10 & HR@5  & NDCG@5 & HR@10  & NDCG@10  \\
\cmidrule{1-13}
Caser   & 0.0116  & 0.0072  & 0.0194 & 0.0097 & 0.0205 & 0.0131 & 0.0347 & 0.0176 & 0.0166 & 0.0107 & 0.0270 & 0.0141  \\
HGN    &  0.0189  & 0.0120  & 0.0313  &  0.0159 & 0.0325  & 0.0206  & 0.0512  & 0.0266  & 0.0321  & 0.0221  & 0.0497  & 0.0277  \\
GRU4Rec   & 0.0129  & 0.0086  & 0.0204  & 0.0110  & 0.0164  & 0.0099  & 0.0283  & 0.0137  & 0.0097  & 0.0059  & 0.0176  & 0.0084 \\
BERT4Rec   & 0.0115   & 0.0075  & 0.0191  &  0.0099 &  0.0203 & 0.0124  & 0.0347  & 0.0170  & 0.0116   & 0.0071  & 0.0203  & 0.0099 \\
FDSA    &  0.0182  & 0.0122  & 0.0288  & 0.0156  & 0.0267  & 0.0163  & 0.0407  & 0.0208  & 0.0228  & 0.0140  & 0.0381  & 0.0189 \\
SASRec &  0.0233  &  0.0154 & 0.0350  &  0.0192 & 0.0387  & 0.0249  & 0.0605  & 0.0318  &  0.0463  &  0.0306  &  0.0675  & 0.0374 \\
S$^3$-Rec & 0.0251  & 0.0161  &  0.0385 & 0.0204 & 0.0387 & 0.0244  & 0.0647  & 0.0327  & 0.0443  & 0.0294  &  0.0700  &   0.0376 \\
P5  &  0.0364   & 0.0296   & 0.0431  & 0.0318  & \bf 0.0508  & \underline{0.0379}  & \underline{0.0664} & \underline{0.0429}  & 0.0608  & 0.0507  & 0.0688  &  0.0534  \\
RecSysLLM  & 0.0360   & 0.0291   & 0.0417  & 0.0302  & \bf 0.0508  & \bf 0.0381  & \bf 0.0667  & \bf 0.0446   &  \bf 0.0676  &  \bf 0.0583  & \bf 0.0712  & \bf 0.0596  \\
\rowcolor{lightthulianpink} P5 {\color{orange}}  & \underline{0.0387}   & \underline{0.0312}  & \underline{0.0460}  &  \underline{0.0336}  &  0.0493  & 0.0367  & 0.0645  & 0.0416  &  0.0587  &  0.0486  & 0.0675  & 0.0536    \\
\rowcolor{lightthulianpink} RecSysLLM {\color{orange}}  & \bf 0.0392   & \bf 0.0330  & \bf 0.0512  &  \bf 0.0375  &  \underline{0.0501}  & 0.0361  &  0.0650 &  0.0407  & \underline{0.0630}  & \underline{0.0523}  & \underline{0.0691}  &  \underline{0.0540}   \\
\bottomrule
\end{tabular}}

\label{tab:sequential}
\end{table*}

\begin{table*}[ht!]
\centering
\caption{Performance on explanation generation (\%). The shadow refers to test on unseen prompts in a zero-shot manner.}

\resizebox{1.8\columnwidth}{!}{
\begin{tabular}{lccccccccccccc}
\toprule
&\multirow{2.5}{*}{Methods} & \multicolumn{4}{c}{\textbf{Sports}} & \multicolumn{4}{c}{\textbf{Beauty}} &  \multicolumn{4}{c}{\textbf{Toys}} \\
\cmidrule(lr){3-6}\cmidrule(lr){7-10}\cmidrule(lr){11-14}
& & BLUE4  & ROUGE1 & ROUGE2  & ROUGEL & BLUE4  & ROUGE1 & ROUGE2  & ROUGEL & BLUE4  & ROUGE1 & ROUGE2  & ROUGEL  \\
\cmidrule{1-14}
 \multirow{4.5}{*}{w/o hints}& Attn2Seq   & 0.5305  & 12.2800  & 1.2107 & 9.1312 & 0.7889 & 12.6590 & 1.6820 & 9.7481 & 1.6238 & 13.2245 & 2.9942 & 10.7398  \\
&NRT   &  0.4793  & 11.0723  & 1.1304  & 7.6674 & 0.8295  & 12.7815  & 1.8543  & 9.9477  & 1.9084  & 13.5231  & 3.6708  & 11.1867  \\
& PETER   & 0.7112  & 12.8944  & 1.3283  & 9.8635  & \underline{1.1541}  & 14.8497  & \underline{2.1413}  & 11.4143  & 1.9861  & 14.2716  & 3.6718  & 11.7010 \\
&P5  &  \underline{1.0407}   &  \underline{14.1589}   & \underline{2.1220}  & \underline{10.6096}  & 0.9742  & \underline{16.4530}  & 1.8858  & \underline{11.8765} & \underline{2.3185}  & \underline{15.3474}  & \underline{3.7209} & \underline{12.1312} \\
&RecSysLLM & \bf 1.2673   & \bf 16.7132   & \bf 2.8980  & \bf 13.0104  & \bf 1.5230  & \bf 19.0032  & \bf 3.0422  & \bf 14.7471 & \bf 2.9923  &  \bf 16.7823  & \bf 4.8372 & \bf 15.0231 \\
\cmidrule{1-14}
\multirow{4.5}{*}{w/ hints} &PETER+  &  2.4627 &  24.1181  & 5.1937  &  18.4105  &  3.2606  & 25.5541  &  5.9668  &  19.7168  &  4.7919  & 28.3083  & 9.4520  &  22.7017 \\
&P5  &  1.4689  &  23.5476  & 5.3926  & 17.5852   & 1.8765  &  25.1183 & 6.0764 & 19.4488  &  3.8933 &  27.9916  & 9.5896 &  22.2178 \\
&RecSysLLM & \underline{3.7232}  & \underline{30.1129}  & \underline{5.0232}  & \underline{20.0020}   & \underline{4.8232} & \underline{26.9832} & \underline{6.2382} & \underline{21.4842}  & \underline{5.9323}  & \underline{29.3232}  & \underline{9.4234} & \underline{23.9843} \\
\rowcolor{lightthulianpink} &P5 &  1.4303  & 23.3810  &  5.3239  & 17.4913   &  1.9031 &  25.1763 & 6.1980  & 19.5188  & 3.5861  & 28.1369  & 9.7562 & 22.3056 \\
\rowcolor{lightthulianpink} &RecSysLLM& \bf  3.9842  & \bf 30.2913  & \bf  5.8923  & \bf 20.3821   & \bf  5.0021 & \bf  27.3854 & \bf 6.7281 & \bf 22.7439  & \bf 6.2912  & \bf 30.2948  & \bf 10.0329 & \bf 24.9932 \\
\bottomrule
\end{tabular}}

\label{tab:explanation}
\end{table*}

\begin{table*}[ht!]
\centering
\caption{Performance on review summarization (\%). The shadow refers to the test on unseen prompts in a zero-shot manner.}

\resizebox{1.9\columnwidth}{!}{
\begin{tabular}{ccccccccccccc}
\toprule
\multirow{2.5}{*}{Methods} & \multicolumn{4}{c}{\textbf{Sports}} & \multicolumn{4}{c}{\textbf{Beauty}} &  \multicolumn{4}{c}{\textbf{Toys}} \\
\cmidrule(lr){2-5}\cmidrule(lr){6-9}\cmidrule(lr){10-13}
 & BLUE2  & ROUGE1 & ROUGE2  & ROUGEL & BLUE2  & ROUGE1 & ROUGE2  & ROUGEL & BLUE2  & ROUGE1 & ROUGE2  & ROUGEL  \\
\cmidrule{1-13}
T0  & 2.1581 & 2.2695  & 0.5694 & 1.6221 & 1.2871 & 1.2750 & 0.3904 & 0.9592 & \underline{2.2296} & 2.4671 & 0.6482 & 1.8424  \\
GPT-2 &  0.7779  &  4.4534  & 1.0033 &  1.9236 &  0.5879 &  3.3844 &  0.6756 &  1.3956 &  0.6221 &  3.7149 &  0.6629 &  1.4813  \\
P5 & \underline{2.6910} &  \underline{12.0314}  &  \underline{3.2921} &   \underline{10.7274} & \underline{1.9325} & \underline{8.2909} & \underline{1.4321}  & \underline{7.4000} & 1.7833 &  \underline{8.7222} & \underline{1.3210} & \underline{7.6134}  \\
RecSysLLM & \bf 4.2823 & \bf 14.8343  & \bf 4.3984 &  \bf 12.4833 &  \bf 3.3821 &  \bf 9.8103 &  \bf 2.8543  &  \bf 10.4003 & \bf 4.0320 &   \bf 12.2932 &  \bf 3.2943 &  \bf 10.4092  \\
\bottomrule
\end{tabular}}

\label{tab:summarize}
\end{table*}

\begin{table*}[ht!]
\centering
\caption{Performance on direct recommendation. The shadow refers to the test on unseen prompts in a zero-shot manner.}

\resizebox{2\columnwidth}{!}{
\begin{tabular}{cccccccccccccccc}
\toprule
\multirow{2.5}{*}{Methods} & \multicolumn{5}{c}{\textbf{Sports}} & \multicolumn{5}{c}{\textbf{Beauty}} &  \multicolumn{5}{c}{\textbf{Toys}} \\
\cmidrule(lr){2-6}\cmidrule(lr){7-11}\cmidrule(lr){12-16}
 & HR@1 & HR@5  & NDCG@5 & HR@10  & NDCG@10 & HR@1 &  HR@5  & NDCG@5 & HR@10  & NDCG@10 & HR@1 &  HR@5  & NDCG@5 & HR@10  & NDCG@10  \\
\cmidrule{1-16}
BPR-MF  & 0.0314   & 0.1404  & 0.0848 &  0.2563 &  0.1220 & 0.0311  & 0.1426  & 0.0857  & \underline{0.2573}  & 0.1224  &  0.0233  & 0.1066  & 0.0641  & 0.2003  &  0.0940 \\
BPR-MLP &  0.0351 & 0.1520  & 0.0927 & 0.2671 & 0.1296  & 0.0317 &  0.1392 &  0.0848 &  0.2542 & 0.1215  & 0.0252  & 0.1142  & 0.0688  & 0.2077  & 0.0988  \\
SimpleX  & 0.0331   &  \bf 0.2362   & \bf 0.1505  &  \underline{0.3290}   &  \underline{0.1800}   &  0.0325  &  \bf 0.2247   &  \bf 0.1441  &  \bf 0.3090  &  \bf 0.1711  & 0.0268   & \bf 0.1958   &  \bf 0.1244   &  \bf 0.2662   & \bf 0.1469  \\
P5   & 0.0641   &  0.1794  & 0.1229  &  0.2598  & 0.1488  & 0.0588 & 0.1573 & 0.1089 & 0.2325 & 0.1330  & \underline{0.0386}   & 0.1122 &  0.0756  &  0.1807 & 0.0975  \\
RecSysLLM  & \underline{0.0654}  &  0.2008  & 0.1438  &  0.2984  & 0.1692  & \bf 0.0618 & \underline{0.1612} & \underline{0.1110} & 0.2209 & 0.1302  &  0.0370  & 0.1301  &  0.0808 &  0.1902  &  0.0998  \\
\rowcolor{lightthulianpink} P5  & 0.0726  &  0.1955  & 0.1355  &  0.2802  & 0.1627  & \underline{0.0608} & 0.1564 & 0.1096 & 0.2300 & \underline{0.1332}  &  \bf 0.0389  & 0.1147  &  0.0767 &  0.1863  &  0.0997  \\
\rowcolor{lightthulianpink} RecSysLLM  & \bf 0.0892  &  \underline{0.2029}  & \underline{0.1502}  &  \underline{0.3001}  & \underline{0.1703}  & 0.6072 & 0.1502 & 0.1097 & 0.2317 & 0.1302  &  0.0327  & \underline{0.1423}  &  \underline{0.0825} &  \underline{0.1926}  &  \underline{0.1028}  \\

\bottomrule
\end{tabular}}

\label{tab:direct}
\end{table*}

We pretrain our RecSysLLM on a diverse set of training examples utilizing different prompt templates across all five task families. This is to thoroughly verify its multitask learning capabilities. The results in Tables \ref{tab:rating}-\ref{tab:direct} demonstrate that for tasks with seen prompt templates, our model reaches the same conclusions as the P5 model and achieves comparable or superior performance. However, we were pleasantly surprised to discover that for unseen prompt templates in a zero-shot manner, our model significantly surpasses P5.

(1) From Table \ref{tab:rating}, for rating prediction, our RecSysLLM gets similar performance on prompt in the train data set, but it has better RMSE and MAE on all three datasets compared with P5 on zero-shot setting. It reflects that our RecSysLLM inherits the semantic understanding capacity of LLM on unseen prompts, which meets our expectations for the LLM.
(2) In Table \ref{tab:sequential}, for the sequential recommendation, our RecSysLLM surpasses P5 on \textit{Beauty} and \textit{Toys}. It gets better performance than P5 on unseen prompts in a zero-shot manner. The results show that our RecSysLLM gains inter- and intra-entity knowledge and make more reasonable predictions.
(3) As shown in Table \ref{tab:explanation}, our RecSysLLM demonstrates superior performance on the task of explanation generation, both with and without feature-based hints. The large improvements in natural language processing abilities of LLMs underlie this strong performance. Moreover, the considerable increase in scores when hints are provided highlights the critical role prompt engineering plays in eliciting the full capabilities of large language models. Through prompt design and the generative power of LLMs, our system achieves state-of-the-art results on this challenging task.
(4) The review summarization results further demonstrate the superiority of our RecSysLLM, as shown in Table \ref{tab:summarize}. Despite having fewer parameters than T0 (7 billion vs 11 billion), our model attains higher performance across all evaluation metrics. These gains over strong baselines like T0 underscore the efficiency and effectiveness of our approach. The capability to produce high-quality summaries with fewer parameters highlights the strength of our method, delivering strong performance without the need for extremely large models.
(5) For the task of direct recommendation, we make an evaluation on open question prompts to test the ability of generative recommendation. The results are illustrated in Table \ref{tab:direct}. Our RecSysLLM outperforms P5 on most evaluation metrics for this task. The simpleX model is a strong collaborative filtering baseline, but RecSysLLM achieves better top-1 item ranking compared to simpleX.

To further analyze the performance gap between the P5 model and our proposed method, we conducted an in-depth examination of the training data. Table \ref{training_example} illustrates that in the P5 model, the items are simply represented by numeric IDs based on their order of occurrence in the dataset. This type of simplistic representation cannot capture semantic information about the items.
In contrast, our RecSysLLM model represents all items as text strings. The textual representation enables our large language model to understand and capture nuanced interrelationships between items much more effectively. We believe this is the primary reason why our model outperformed P5 across most cases. The textual representation in our model empowers it to ingest semantic details and identify meaningful connections that cannot be derived from IDs alone.

\section{Applications in real-world dataset}

\subsection{Dataset}
The data used in this work was collected from Alipay, a mobile payment platform in China. We extracted user behavior logs, including bills, search queries, and page visits for several recommendation tasks. Each user sequence consists of the user's 500 most recent interactions, spanning over one year of history for some users. The user sequences are used to model evolving user interests and capture both long- and short-term preferences. The training set contains $200,000$ sequences, and the test set contains $10,000$ sequences. The large-scale real-world dataset enables the modeling of complex user behavior and preferences for various recommendation tasks. The hierarchical categories and sequential interactions provide rich signals for understanding user interests.

\subsection{Implementation Details}
Our RecSysLLM model for Chinese language tasks leverages the powerful ChatGLM-6B \cite{du2021glm} model as a foundation. ChatGLM-6B is an open-source bilingual language model with 6.2 billion parameters, trained on a trillion-token corpus comprised primarily of Chinese text with some English. The model architecture is based on the General Language Model (GLM) framework. Similarly, our approach builds on this pre-trained ChatGLM-6B foundation by utilizing LoRA to adapt the model to our specific recommender system tasks. We set the rank of Lora to $8$, which is a proper coefficient chosen by the ablation study.

\subsection{Sequential Recommendation.}
\paragraph{Task Description.} In this section, we conduct two sequential recommendation tasks to evaluate the performance of our model, i.e., next-item prediction and candidate recommendation. For next-item prediction, the model directly predicts the next item a user will interact with based on their historical interactions and profiles. For candidate recommendation, given a user's interaction history, profiles, and a list of candidate items where only one is positive, the model chooses the correct next item. We have benchmarked our model on the Amazon Sports, Beauty, and Toys datasets and demonstrated superior recommendation capabilities compared to other baseline recommender systems. Here, we compare our RecSysLLM to the powerful generative models ChatGPT and the recently announced GPT-4. We also compare our method against a basic fine-tuning approach of ChatGLM on our recommendation tasks. This allows us to analyze the improvements gained by our specialized techniques that are tailored for the recommendation systems based on LLM. By evaluating against a simple fine-tuning baseline, we can quantify the benefits of our proposed approach and demonstrate that our architectural choices and training methodology confer meaningful advantages on recommendation performance compared to just fine-tuning a large language model out-of-the-box. 

\paragraph{Next Item Prediction.} The results in Table \ref{tab:NextItem_result} demonstrate that for next-item prediction, our RecSysLLM achieves performance on par with ChatGPT, with both significantly outperforming the naive ChatGLM fine-tuning and GPT-4. This is a surprising result, as we expected the larger GPT-4 model to achieve superior performance compared to ChatGPT on this recommendation task due to its greater parameter size and pretraining scale. However, GPT-4 did not exhibit particularly strong results and was not decisively superior to ChatGPT. There are several potential explanations for why GPT-4 underperformed expectations on the next item prediction. First, the dataset and evaluation methodology used for this task may not have fully exercised GPT-4's strengths in areas like few-shot learning and knowledge recall. Second, GPT-4's more powerful generative capabilities may have caused it to diverge too far from the tight distributions of the recommendation data. There could be a mismatch between GPT-4's broad natural language generation skills and the specialized prediction required by the recommender system task. In summary, our specialized RecSysLLM demonstrates that simply utilizing a larger pre-trained language model is not the only path to improved recommendation performance. The model architecture and pretraining objectives also play a vital role. By designing a model specifically for the recommendation, focusing the pretraining on recommendation data, and tightly bounding the final fine-tuning, our RecSysLLM is able to match or exceed the performance of even much larger general language models like GPT-4 for next-item prediction. These results highlight the importance of specialized model design in addition to scale for advancing recommendation systems.

\paragraph{Candidate Recommendation.} 
For candidate recommendation in Table \ref{tab:Candidate_result}, our RecSysLLM consistently outperforms both ChatGPT and the naive ChatGLM fine-tuning across metrics. This demonstrates the effectiveness of our specialized approach for this task. In contrast to the next item results, this time, GPT-4 achieves the overall best performance on candidate recommendation. In candidate recommendation, given a user's interaction history, profile, and a list of candidate items where only one is the ground truth next interaction, the model must choose the correct item from the candidates. With a constrained set of options provided, GPT-4 is able to give full play to its powerful reasoning and deduction capabilities. The limited choice set prevents GPT-4's generative tendencies from leading it astray. As a result, GPT-4 is able to leverage its scale and pretraining to achieve the best overall performance on candidate recommendation. In summary, by providing GPT-4 a focused set of candidates, we can elicit its strengths in logical reasoning while avoiding over-generation. This allows GPT-4 to achieve state-of-the-art results on candidate recommendation, showcasing the benefits of its scale and pretraining. Our specialized RecSysLLM still exceeds the general language models on this task, demonstrating the value of recommendation-specific modeling. But these results highlight how large generative LMs like GPT-4 can excel given the right setup. 

\begin{table}[ht!]
\centering

\caption{Performance on next item recommendation.}

\resizebox{0.82\columnwidth}{!}{
\begin{tabular}{l|cccc}
\toprule

Methods & HR@5  & NDCG@5 & HR@10  & NDCG@10  \\
\midrule

ChatGPT  & \bf 0.4326 & \bf 0.3208  & \bf 0.5110 & 0.3465 \\
GPT-4 &   \underline{0.3846} &   0.2890 &  0.4674 &   0.3159 \\
ChatGLM+SFT   & 0.2654  & 0.2091  & 0.3729 &  0.2513 \\
RecSysLLM &  0.3805 & \underline{0.3072}  & \underline{0.4756} &  \bf 0.4091  \\
\bottomrule
\end{tabular}}
\label{tab:NextItem_result}
\end{table}

\begin{table}[ht!]
\centering

\caption{Performance on candidate recommendation task.}

\resizebox{0.9\columnwidth}{!}{
\begin{tabular}{l|ccccc}
\toprule

Methods & HR@1 & HR@5  & NDCG@5 & HR@10  & NDCG@10  \\
\midrule

ChatGPT  &  0.3786 & 0.5550  & 0.4715 & 0.6424 & 0.5001 \\
GPT-4 &  \bf 0.7079 & \bf 0.8154  & \bf 0.7671 & \bf 0.8560 & \bf 0.7804 \\
ChatGLM+SFT   &  0.2984 &  0.7012 & 0.6826 & 0.7621 & 0.7038  \\
RecSysLLM & \underline{0.4965}  & \underline{0.7435}  & \underline{0.7032} &  \underline{0.7728}  & \underline{0.7237}\\
\bottomrule
\end{tabular}}

\label{tab:Candidate_result}
\end{table}

\section{Conclusion}
The focus of this paper is to design a novel paradigm of pretraining recommendation models based on large language models. We introduce a novel mask mechanism, span order, and positional encoding to inject inter- and intra-entity knowledge into the LLM. Although our method follows the architecture of generative language models (GLM) to some extent, the core ideas of special designs for entities in recommendation tasks can be extended to other large language models. The experiments conducted on public and industrial datasets demonstrate the effectiveness and potential of our proposed model on recommendation systems and related applications. The results show improvements over strong baselines, indicating that encoding entity relationships during pretraining can meaningfully improve downstream performance. While we validate our approach on a select set of datasets, further experiments on a wider range of tasks would better reveal the strengths and limitations of the method. In particular, evaluating the approach across a more diverse set of domains could shed light on how robust the learned representations are. Additionally, from the perspective of causal inference \cite{yao2021survey,chu2023causal}, there are likely further improvements to be made in terms of how semantic connections between entities are captured and injected into the model.

\bibliography{aaai24}

\clearpage
\appendix

\section{Recommendations and LLM}

\subsection{Motivation}

Compared with recommendation models based on large language models (LLMs), conventional recommendation models \cite{hidasi2015session,tang2018personalized,kang2018self,sun2019bert4rec,geng2022recommendation} trained from scratch using architectures like Transformer \cite{vaswani2017attention}, Bert \cite{devlin2018bert}, RNN \cite{schuster1997bidirectional}, CNN \cite{krizhevsky2012imagenet} have several key limitations. First, they lack a deep understanding of context and semantics that comes from pretraining a large model on diverse corpora. As a result, they struggle to truly comprehend user preferences and behavioral sequences. Second, they have minimal ability to generate novel, high-quality recommendations since they are not optimized for free-form text generation. LLMs, in contrast, can produce human-like recommendations by leveraging their generative capabilities. Third, conventional models have difficulty effectively leveraging multiple data modalities like text, images, audio, etc. LLMs are adept at multimodal processing due to pretraining objectives that learn connections between modalities. Finally, LLMs can seamlessly adapt to new downstream recommendation tasks through simple fine-tuning, whereas conventional models require extensive retraining. For example, BERT4Rec \cite{sun2019bert4rec} employs deep bidirectional self-attention to model user behavior sequences. They are trained solely based on the recommendation data without the general knowledge corpus, resulting in a limited understanding and reasoning of behavior sequence data and an inability to empower downstream tasks better. In summary, recommendation models based on pretrained LLMs are more contextual, creative, versatile, and adaptable compared to conventional models trained from scratch.

\subsection{Current Development}

Although the application of LLMs like ChatGPT in recommendation has not been widely explored yet, some novel investigations have emerged recently that show their promising potential in this domain. There are mainly three categories.

(1) \textit{LLM as a recommendation system.} First, Unlike traditional recommendation methods, they do not retrain a new model, relying only on the prompts of LLM \cite{liu2023chatgpt,gao2023chat,dai2023uncovering,chen2023palr} or slight fine-tuning \cite{zhang2023recommendation,kang2023llms,bao2023tallrec} to convert recommendation tasks into natural language tasks. 
They always design a set of prompts on recommendation scenarios, including rating prediction, sequential recommendation, direct recommendation, explanation generation, and review summarization. They explore the use of few-shot prompting to inject interaction information that contains user potential interest to help LLM better understand user needs and interests. 

(2) \textit{LLM as supplementary information via embeddings or tokens.} This modeling paradigm \cite{wu2021empowering,qiu2021u,yao2022reprbert,muhamed2021ctr,xiao2022training} views
the language model as a feature extractor, which feeds the
features of items and users into LLMs and outputs corresponding embeddings. A traditional RS model can utilize
knowledge-aware embeddings for various recommendation tasks. This approach \cite{liu2023first,wang2022towards,wang2023generative} generates tokens based on the inputted items’
and users’ features. The generated tokens capture potential preferences through semantic mining, which can be
integrated into the decision-making process of a recommendation system.

(3) \textit{LLM as Agent.} As an agent, the large model assists in scheduling the entire recommendation model for recommendations and is responsible for pipeline control. Specifically, these models \cite{andreas2022language,bao2023tallrec,hou2023large,lin2023can,gao2023chat,friedman2023leveraging} help to adapt LLM to the recommendation domain, coordinate user data collection, feature engineering, feature encoder, scoring/ranking function.

\subsection{Challenges}
Compared to superficially leveraging large language models, our purpose is built on the large language model, maximizing the preservation of knowledge and logical reasoning abilities from the original large language model to ensure the inference for the behavioral sequences and fluent generation of downstream sub-tasks, while also achieving the recommendation function by learning user profile features and user behavior sequences. The crucial aspect of harnessing the power of language models in enhancing recommendation quality is the utilization of their high-quality representations of textual features and their extensive coverage of external knowledge to establish correlations between items and users. \cite{wu2023survey}. Therefore, we need to preserve the tokenization, parameters, and architecture of the large language model as much as possible. For example, Pretrain, Personalized Prompt, and Predict Paradigm (P5) \cite{geng2022recommendation} is established upon a basic encoder–decoder framework with Transformer blocks to build both the encoder and decoder. Although it is built on T5 \cite{raffel2020exploring}, it modified the structure of the model by adding additional positional encodings and whole-word embeddings, which will partially destroy the original knowledge in the language model. 

Notably, there is a difference in the format of the data. Large language models are trained on vast amounts of logically structured text, with consistent reasoning, logical thought processes, and proper grammar. In contrast, recommendation systems analyze digital user features, fixed item entities, and incoherent behavioral sequences. Additionally, The purpose of training data for large language models is to teach the model how to understand language and generate new text that is similar to the training data. Conversely, the purpose of user behavioral sequence data in recommendation systems is to dig a deeper understanding of user preferences, behavior sequences, and relationships between them so that to provide personalized recommendations. 

Therefore, building a recommendation system on top of a large language model that retains the LLM's knowledge and logical reasoning abilities, while also achieving the recommendation function by learning user profile features and user behavior sequences poses significant challenges.

\section{Baselines in Benchmark Experiments}

To showcase our competence in a wide range of recommendation-related tasks, we employ representative approaches for different tasks, including Rating Prediction, Direct Recommendation, Sequential Recommendation, Explanation Generation, and Review Summarization, that have been previously used by \cite{geng2022recommendation}. The summary of baseline methods for five different task families is provided in Table \ref{tab:baseline}.

\noindent\textbf{Rating Prediction.~} 
This task involves incorporating user-item rating data as part of the training set, where item ratings are represented numerically. The model is asked questions with prompts, and it outputs corresponding rating values.
The baselines for this task are \textbf{MF}~\cite{koren2009matrix} and \textbf{MLP}~\cite{cheng2016widedeep}.

\noindent\textbf{Direct Recommendation.} 
For direct recommendation, we employ classic algorithms \textbf{BPR-MF} \cite{rendle2009bpr}, \textbf{BPR-MLP} \cite{cheng2016widedeep} and \textbf{SimpleX} \cite{mao2021simplex} as baselines. They showcase the effectiveness of direct recommendation tasks when utilizing non-semantic information as features. This allows us to gain a more comprehensive understanding of the potential of recommendations given by LLM-based models.

\noindent\textbf{Sequential Recommendation.} The sequential recommendation task utilizes the user's historical interaction sequences as input to predict the next item.
We compare our proposed approaches with representative baselines in the field.
Among that, some models aim to model the Markov Chain of user interactions by way of neural network architectures like convolutional neural networks, recurrent neural networks, and attention-based modules. \textbf{Caser}~\cite{tang2018personalized}  employs convolutional neural networks to model user interests. \textbf{HGN}~\cite{ma2019hierarchical} adopts hierarchical gating networks to capture user behaviors from both long and short-term perspectives. \textbf{GRU4Rec}~\cite{hidasi2016gru4rec}  utilizes recurrent neural network to model the user click history sequence. \textbf{SASRec}~\cite{kang2018self} and \textbf{FDSA}~\cite{zhang2019feature} use self-attention modules to model feature transition patterns for sequential recommendation and the former combine RNN-based approaches to retain the sequential properties of items.
\textbf{BERT4Rec}~\cite{sun2019bert4rec} adopts the BERT-style masked language modeling to learn the relations among items from the perspective of bidirectional representations in the recommendation. It started to use methods in neural language processing, but BERT did not have a strong semantic understanding capacity at that time. 
\textbf{S$^3$-Rec}~\cite{zhou2020s3} leverages self-supervised objectives to enhance the discovery of correlations among different items and their attributes.

\noindent\textbf{Explanation Generation.} We evaluate the task of  explanation generation by comparing the performance of several baseline models. \textbf{Attn2Seq}~\cite{EACL17-Att2Seq} and \textbf{NRT}~\cite{li2017neural} utilizes the neural network to encode attributes of user and item into vectors and then invokes an attention mechanism or GRU~\cite{cho2014learning} to generate reviews conditioned on the attribute vector.
\textbf{PETER}~\cite{li2021personalized} use Transformer architecture and designa a modified attention mask. The variant \textbf{PETER+} takes a hint feature word to augment the process of generating explanations.

\noindent\textbf{Review Related.} For review summarization, we adopt pretrained \textbf{T0}~\cite{sanh2022multitask} and \textbf{GPT-2}~\cite{radford2019language} as baselines. The  latter model parameters were obtained from Hugging Face\footnote{\url{https://huggingface.co/}}, which is a big platform to share models, datasets, and applications.

\begin{table}[t!]
\centering
\small
\caption{The summary of baseline methods for five different task families.}
\resizebox{1\columnwidth}{!}{
\begin{tabular}{l|l l}
\toprule
Rating Pre    &  \textbf{MF}~\cite{koren2009matrix}&  \textbf{MLP}~\cite{cheng2016widedeep}      \\
\midrule
\multirow{2}{*}{Direct Rec}    &  \textbf{BPR-MF} \cite{rendle2009bpr} &\textbf{BPR-MLP} \cite{cheng2016widedeep}    \\

 & \textbf{SimpleX} \cite{mao2021simplex}&\\
\midrule
\multirow{4}{*}{Sequential Rec} & \textbf{Caser}~\cite{tang2018personalized} &\textbf{HGN}~\cite{ma2019hierarchical} \\

&\textbf{GRU4Rec}~\cite{hidasi2016gru4rec}& \textbf{BERT4Rec}~\cite{sun2019bert4rec}\\

& \textbf{FDSA}~\cite{zhang2019feature}&\textbf{SASRec}~\cite{kang2018self} \\
&\textbf{S$^3$-Rec}~\cite{zhou2020s3}
&\textbf{BERT4Rec}~\cite{sun2019bert4rec} \\

\midrule
\multirow{2}{*}{Explanation Gen} & \textbf{Attn2Seq}~\cite{EACL17-Att2Seq} &
\textbf{NRT}~\cite{li2017neural} \\
& \textbf{PETER}~\cite{li2021personalized}&  \textbf{PETER+} \\
\midrule
Review Sum & \textbf{T0}~\cite{sanh2022multitask} &\textbf{GPT-2}~\cite{radford2019language} \\
\bottomrule
\end{tabular}}
\label{tab:baseline}
\end{table}

\section{Further Analysis in the real-world dataset} 

\begin{figure*}
\centering
\includegraphics[width=2\columnwidth]{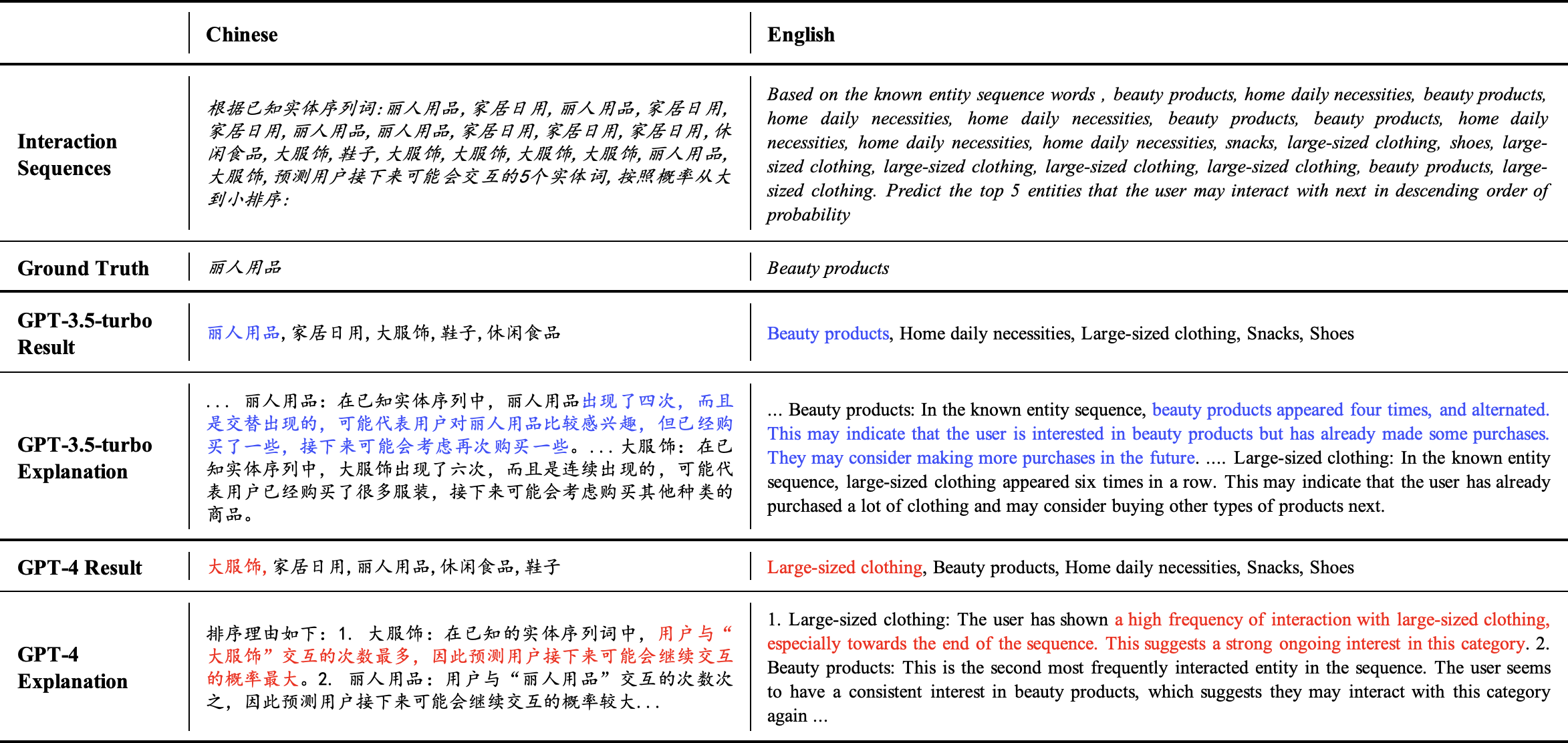}
\caption{The case studies of ChatGPT and GPT-4 for next item recommendation in the real-world dataset.}
\label{fig:case_analysis}
\end{figure*}

In addition to optimizing the recommendation performance, it is also important to understand why large language models like ChatGPT and GPT-4 are able to effectively conduct recommendation tasks in the first place. To explore this further, we provide several real-world case studies in Figure \ref{fig:case_analysis}, where we systematically probe and dissect the reasoning process of these models when making recommendations, using carefully designed prompt-based queries. This analysis sheds light on the strengths and limitations of relying solely on the knowledge and reasoning capabilities embedded in large pre-trained language models for recommendation tasks, and points towards potential areas for improvement.

\begin{figure}
\centering
\includegraphics[width=0.9\columnwidth]{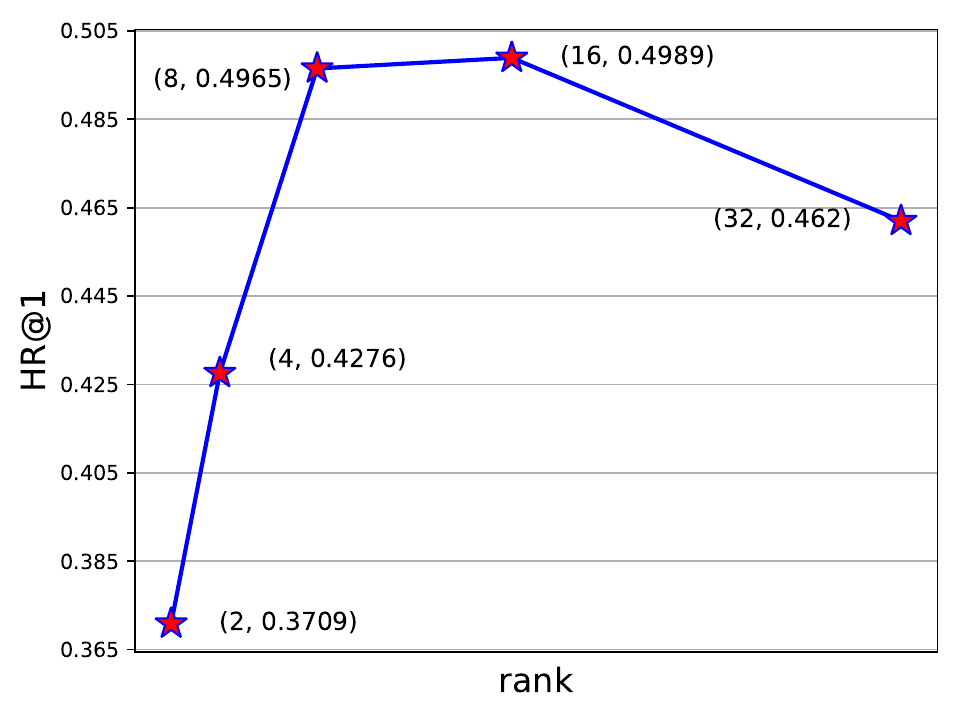}
\caption{The HR@1 with different rank $r$ of LoRA.}
\label{fig:rank}
\end{figure}

Our experiments also analyze the impact of the rank $r$ of Low-Rank Adaptation on model performance. We evaluate five different rank values - $2, 4, 8, 16,$ and $32$ - to determine the optimal balance between model capacity and predictive ability. As shown in Figure \ref{fig:rank}, we find that a rank of 8 provides sufficient learning capacity, with minimal improvements from increasing to 16. This indicates that capturing inter- and intra-entity relationships requires only a small number of additional trainable parameters beyond the base LLM, without the need for substantial model expansion. Rank 8 strikes the right balance, enabling Low-Rank Adaptation to boost performance through targeted parameterization rather than sheer scale. Overall, our results demonstrate that Low-Rank Adaptation offers an efficient approach to entity-aware language modeling.

\end{document}